\documentclass[twocolumn,showpacs,preprintnumbers,amsmath,amssymb]{revtex4}

\usepackage{graphicx}% Include figure files
\usepackage{dcolumn}% Align table columns on decimal point
\usepackage{bm}% bold math

\usepackage{amsmath}
\usepackage{url}

\begin{document}

\title{Neutrino Pair Emission from Hot Nuclei During Stellar Collapse}

\author{G. Wendell Misch$^{1}$}
\author{B. Alex Brown$^{2}$}
\author{George M. Fuller$^{1}$}
\affiliation{$^{1}$Department of Physics, University of California, San Diego, La Jolla, CA 92093, USA}
\affiliation{$^{2}$ National Superconducting Cyclotron Laboratory, and Department of Physics and Astronomy, Michigan State University, East Lansing, Michigan 48824, USA}

\date{\today}

\begin{abstract}
We present shell-model calculations showing that residual interaction-induced configuration mixing enhances the rate of neutral current de-excitation of thermally excited nuclei into neutrino-antineutrino pairs.  Though our calculations reinforce the conclusions of previous studies that this process is the dominant source of neutrino pairs near the onset of neutrino trapping during stellar collapse, our shell-model result has the effect of increasing the energy of these pairs, possibly altering their role in entropy transport in supernovae.
\end{abstract}

\maketitle

\section{Introduction}
In this paper we study the role of the nucleon-nucleon interaction in the process of de-excitation of hot, excited nuclei into virtual $Z^0$'s and
neutrino-antineutrino pairs. This process is likely the dominant source of neutrino pairs in collapsing stellar cores \cite{fm:1991,km:1979,Horowitz:1992lr}. The energy of the neutrinos in these pairs, set in part by nuclear structure considerations, can be an important determinant of entropy transport in core collapse supernovae. In turn, the entropy figures prominently in the nuclear composition, neutronization history, and initial shock energy in supernova models.

The entropy per baryon in the collapsing stellar core is low and, as a result, most nucleons reside in large nuclei and there are very few free protons \cite{bbal:1979,fuller:1982}.  The paucity of free protons has the effect of suppressing the overall electron capture rate \cite{fuller:1982}, yielding a greater electron fraction $Y_{e}$ (electrons per baryon).  Most pressure support within the core comes from electron degeneracy, so higher $Y_e$ during collapse implies a larger pressure and, hence, a larger homologous core mass.  The mass of this inner core determines the initial energy of the post-bounce shock: a more massive inner core yields a stronger initial shock.  The strength of the shock, the mass of the core above the shock, and photo-dissociation of heavy nuclei in this outer core (all determined in part by $Y_{e}$) are important parameters in the supernova explosion process \cite{arnett:1977, bw:1982,bw:1985, bmd:2003, bm:2007, sjfk:2008, bbol:2011,ajs:2007,Liebendorfer:2008uq,Liebendorfer:2009qy,hjm:2010,Hix:2010fk,Burrows:2012lr}.

In the epoch near neutrino trapping, when the core density is $\sim{10}^{12}\,{\rm g}\,{\rm cm}^{-3}$, the electron fraction is $Y_e \approx  0.32$ \cite{arnett:1977}, giving an electron Fermi energy $\mu_e \approx 51.5\,{\rm MeV}\,\left( \rho_{12}\, Y_e \right)^{1/3} \approx 35\,{\rm MeV}$, where the density is scaled as $\rho_{12} \equiv \rho/{10}^{12}\,{\rm g}\,{\rm cm}^{-3}$. The temperature of the core is in the neighborhood of $T\sim \,1\,{\rm MeV}$ to $2\,{\rm  MeV}$, so the electrons are highly degenerate.  (In this paper we use natural units and set $\hbar = k_{\rm b} = c=1$.) Energy emission from the core via neutrinos helps to maintain low entropy, but at a core density of $\rho_{12}\sim 1$, high energy neutrinos are trapped by neutral current coherent scattering on nuclei.  However, since the cross-section for this process varies as the square of neutrino energy, low energy ($E_\nu < 10\,{\rm MeV}$) neutrinos may escape, carrying away entropy and possibly, depending on the process, lepton number.

There are a number of ways to produce low energy neutrinos in the core: inelastic down-scatter of neutrinos on electrons \cite{Tubbs:1975lr,Tubbs:1979fk}; electron neutrino-pair bremsstrahlung; plasmon decay, {\it etc.} \cite{Beaudet:1967rt,Barkat:1975yq,dkst:1976,Itoh:1983fj,Schinder:1987kx,Itoh:1989uq,Kohyama:1993qy,Itoh:1996fk,Itoh:1996lr}.  However, since these processes involve electron interactions, most of them are suppressed by the extreme electron degeneracy. In these conditions there simply isn't much phase space available for electron down-scatter.

This leads us to explore low-energy neutrino production mechanisms that do not involve electrons.  Simple neutral current neutrino-nucleon down-scattering tends to be ineffective in this regard. This is because this process is roughly conservative: the nucleon mass is large compared to the typical electron capture-generated neutrino energies, $E_\nu \sim 25\,{\rm MeV}$. Free nucleon neutrino-pair bremsstrahlung \cite{Flowers:1975lr,Reddy:1998fk,Prakash:2001qy}, a key process in neutron star cooling, is less effective here because of phase space considerations and because there are few free nucleons in the low entropy conditions that favor large nuclei during stellar collapse.

Indeed, there are analogs of these two processes for nucleons that reside in the large nuclei characterizing the neutrino trapping epoch, and these are not subject to the limitations of their free nucleon cousins.  These analogs are: inelastic down-scattering of energetic neutrinos on nuclei; and neutrino pair emission from thermally-excited nuclei \cite{btz:1974,km:1979,fm:1991}.  The first of these processes has a relatively large cross section, as this channel is what Ref.~\cite{fm:1991} termed an \lq\lq up\rq\rq\ transition, where the nucleus acquires energy from the neutrino, implying that nucleons transition to higher nuclear excited states, where they are relatively  less Pauli blocked. These processes can have
important implications for supernovae \cite{Bruenn:1991uq,fm:1991,Sampaio:2002qy,Martinez-Pinedo:2008fk,Langanke:2008lr}.

\begin{figure}[here]
\centering
\includegraphics[scale=.7]{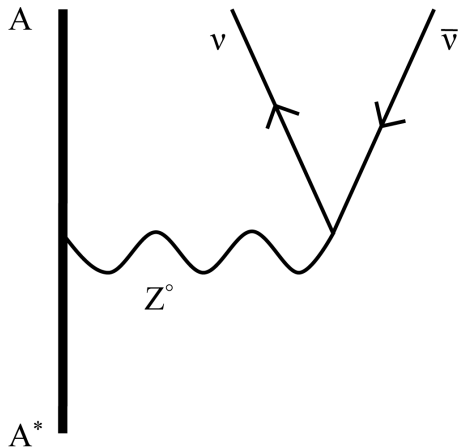}
\caption{Neutral current neutrino pair emission from an excited nucleus A*.}
\label{fig:feyn}
\end{figure}

In contrast, the second of these, the nuclear de-excitation into neutrino pairs channel shown in Fig.~\ref{fig:feyn}, is a \lq\lq down\rq\rq\ transition,
subject to more nuclear Pauli blocking, and therefore possessing considerably less nuclear weak interaction strength on average than the neutrino inelastic down-scatter channel. Nevertheless, the de-excitation process has some unique features: in principle it may produce lower energy neutrinos than the down-scattering channel and, should these escape the core, entropy will be lost but electron lepton number will not be. As shown in Ref.~\cite{Horowitz:1992lr}, hot nuclei can also de-excite into neutrino pairs through a virtual plasmon (photon collective mode in the plasma), and this process has been argued to lead to large enhancement factors in nuclear neutrino pair emission in the first forbidden channel. As we will show, our nuclear structure considerations also impact this channel. All of these issues depend to some extent on the nuclear physics of down transitions, and so this is where we concentrate in this work. 

\begin{figure}[here]
\centering
\includegraphics[scale=.7]{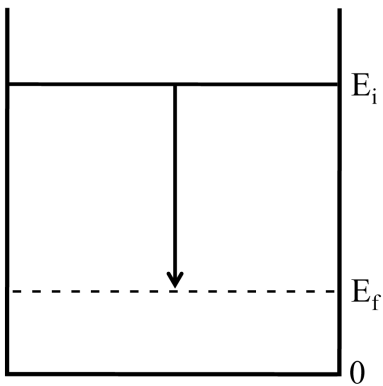}
\caption{Thermally populated nuclear state with excitation energy E$_{i}$ de-excites via virtual Z$^{0}$ emission to a final state with excitation energy E$_{f}$.}
\label{fig:ei-ef}
\end{figure}

In Section II we discuss the nuclear and phase space aspects of de-excitation into neutrino pairs.  Nuclear shell-model considerations are discussed in section III. In Section IV we discuss results, and in Section V we give conclusions.

\section{De-Excitation Rates and Nuclear Structure}

\subsection{Large, Highly-Excited Nuclei During Core Collapse} 

Strong and electromagnetic interactions are fast enough that the material in stellar collapse can be in thermal and chemical equilibrium, {\it i.e.,} Nuclear Statistical Equilibrium (NSE). The weak interaction also is driving toward equilibrium (beta equilibrium) at the epoch of neutrino trapping, but has not yet arrived there. The pioneering work by Bethe {\it et al.}, Ref.~\cite{bbal:1979}, showed that the entropy-per-baryon in a collapsing iron core is $s \approx 1$ and, as a consequence, most all nucleons will reside in large nuclei. Minimizing the free energy for typical conditions, for example with $\rho_{12} \sim 1$ and $T = 1\,{\rm MeV}$ to $2\,{\rm MeV}$, yields a mean nuclear mass $A \sim 100$. 

The mix of nuclei in NSE in these conditions is exotic. These huge nuclei will have fair neutron excess, because $Y_e < 0.4$, implying neutron-to-proton ratios $n/p > 1.5$. Moreover, because the nuclear level density is high, these nuclei will be sitting at high excitation energies. To see this we can treat the nucleons in the nucleus as a Fermi gas. Using the familiar Bethe approximation \cite{bethe:1936} for the nuclear level density, an estimate of the average nuclear excitation energy is \cite{bbal:1979} 
\begin{equation}
\label{eq:bethe}
\langle E \rangle \approx a\, T^2,
\end{equation}
where the level density parameter is $a\approx A/8$ MeV$^{-1}$.  For example, with $A =100$ and $T = 2\,{\rm MeV}$, Eq.~\ref{eq:bethe} implies an excitation energy $\langle E\rangle \sim 50\, {\rm MeV}$. The expression in Eq.~\ref{eq:bethe} is easily understood: The number of nucleons excited above the nuclear neutron and proton Fermi levels will be $\approx a\, T$, and in thermal equilibrium each nucleon so excited will have an average excitation $\sim T$.

\subsection{Nuclear De-Excitation Rates}

Consider an excited nucleus dropping down to a lower excitation energy via virtual $Z^0$ emission, as depicted in Fig.~\ref{fig:ei-ef}. This is similar to a nuclear M1 gamma transition. The de-excitation rate \cite{btz:1974,fm:1991} from initial nuclear state $\vert i \rangle$, with excitation energy $E_i$, to final nuclear state $\vert f \rangle$, with excitation energy $E_f$, is
\begin{eqnarray}
\label{eq:fm}
\lambda_{if} & \approx &  \frac{G_{\rm F}^2\, g_{\rm A}^2}{60\, \pi^3}\,{\left(\Delta E\right)}^5\,B(GT)_{if}
\\
\label{eq:btz}
& \approx & 1.71 \times 10^{-4}\,{\rm s}^{-1}\, \left(\frac{\Delta E}{\text{MeV}} \right)^{5}B(GT)_{if}
\end{eqnarray}
where $G_{\rm F}$ is the Fermi constant, $g_{\rm A}\approx 1.26$ is the axial vector coupling constant, $\Delta E=\vert E_i-E_f\vert$ (hereafter referred to as the transition energy) is the difference between the initial and final state nuclear excitation energies, and $B(GT)_{if} = \vert \langle f\vert \vert \Sigma_{k} (\overrightarrow{\sigma}\ t_z)_{k} \vert \vert i \rangle \vert ^{2}/(2J_{i}+1)$ is the reduced transition probability associted with the axial vector operator. The matrix element connects initial nuclear state $\vert i\rangle$ with final nuclear state $\vert f\rangle$. Here $\overrightarrow{\sigma}$ is the Pauli operator and $t_z$ is the $z$-component of nuclear isospin. For the nuclei we consider here, only the axial vector matrix element is significant: when we neglect the relatively small Coulomb effects, the nuclear part of the Hamiltonian commutes with isospin operators ({\it e.g.,} $T_z$), and the weak vector matrix element $\vert \langle f \vert T_z \vert i \rangle \vert ^{2}$ is zero.

The corresponding neutrino-plus-antineutrino energy emission rate, $\Lambda_{if}$, for this transition is the product of the de-excitation rate and the transition energy.  Whether or not the neutrinos carrying this energy escape from the star without scattering, thereby turning the energy emission rate into an {\it energy loss rate}, depends on many factors, most especially the neutrino energies. 

The total energy emission rate for an excited nucleus in initial state $\vert i\rangle$ is the sum of the energy emission rates to all accessible final states, and can be viewed as a function of $E_i$,

\begin{equation}
\Lambda_i^{\rm tot}\left( E_i\right) = \sum_{f,\ E_f \leq E_i} {\vert E_f-E_i
\vert\, \lambda_{if}}.
\label{eq:ratetotal}
\end{equation}

The total overall energy emission rate for the entire nucleus follows on performing a population index-weighted sum over all initial states $i$,

\begin{equation}
\Lambda_{\rm tot} =  \sum_i{ P_i\ \Lambda^{\rm tot}_i } \approx
{\frac{1}{Z}}\,\int_0^\infty{{\tilde{\rho}\left( E_i \right)}\,
e^{-E_i/T}\,\Lambda_i^{\rm tot} \,dE_i},
\label{eq:total}
\end{equation}

where $P_i = (2 J_i + 1) \exp(-E_i/T)/Z$ is the population index for state $i$, with $J_i$ the spin of level $i$, $Z=\sum_i{ (2 J_i + 1) \exp(-E_i/T)}$ is the nuclear partition function, and $\tilde{\rho}(E_i)$ is the nuclear level density at excitation energy $E_i$.

For the thermodynamic conditions relevant for NSE near $\rho_{12} \sim 1$, we can get a crude estimate of $\Lambda_{\rm tot}$ by simply evaluating $\Lambda_i^{\rm tot}$ at the mean excitation energy for temperature $T$, {\it i.e.,} taking $E_i =\langle E\rangle$,

\begin{equation}
\Lambda_{\rm tot} \approx \Lambda_i^{\rm tot}\left( E_i =\langle E\rangle
\right).
\label{eq:approx}
\end{equation}

The rationale for this approximation is that while the nuclear level density rises nearly exponentially with excitation energy, the Boltzmann factor in Eq.~\ref{eq:total} falls exponentially with this energy, so that their product is strongly peaked at $\langle E\rangle$. 

However, using this rough approximation is problematic. The level density is high near $\langle E\rangle$, and there will be many different kinds of nuclear many-body states with {\it e.g.,} different spins and isospins, but all with roughly this excitation energy. Therefore, choosing a single representative state is not possible. 

\section{Shell-Model Considerations}

\subsection{Approaches to the Problem}

Evaluating the energy emission rates in Eq.~\ref{eq:ratetotal} and Eq.~\ref{eq:total} for a nucleus with nuclear mass number $A\sim 100$ at a mean excitation energy $\sim 50\,{\rm MeV}$ is clearly impractical with conventional nuclear structure techniques tailored to capture low excitation energy physics. The shear size of the problem, some two dozen particles excited above the Fermi surface in a mass $\sim 100$ nucleus with all of the fp-, gd-, and gh-shells in play, precludes this route.

There are two possible alternative approaches: (1) Treat the $\sim a\,T$ nucleons excited above the Fermi sea as nearly free particles within a dense environment, with appropriate phase space modifications, and then calculate the neutrino-pair bremsstrahlung rates for these; and (2) Exploit the fact that each nucleon excited above the nuclear Fermi level has only a relatively small amount of energy ($\sim T$), so that conventional shell-model treatments are efficacious, at least for nuclei with {\it low enough mass} that the problem is tractable computationally. The first of these, by treating valence nucleons as plane waves \cite{km:1979}, will tend to overestimate \cite{fm:1991} the nuclear weak strength available, but has the advantage that it would go smoothly to the homogeneous matter limit when nuclei merge at high density ($\rho_{12} > 10$). 

Here we will take up the second approach, in part because it has the advantage of getting a better handle on the weak nuclear strength, nuclear structure effects, and energetics. The latter is an especially critical issue since the neutrino energy emission rates scale like six powers of the transition energies. Therefore, ascertaining how {\it e.g.,} configuration mixing and particle-hole repulsion act is important. However, we will have to model nuclei with lower masses, generally sd-shell and fp-shell species, rather than the mass $\sim 100$ nuclei that neutrino trapping NSE conditions pick out. At best, this approach will allow us to see trends that may at some point facilitate extrapolation of these considerations to enable estimates for rates from heavier nuclei.  

\subsection{Extension to Heavy Nuclei}

Even calculating the de-excitation rate from a single level in a \lq\lq small\rq\rq\ nucleus is a challenging and unusual nuclear structure problem, in part because matrix elements between highly excited states are required.  We have approached this using a conventional nuclear shell model with the usual filled closed core of nucleons in low-lying single-particle states plus valence nucleons in a model space. We then employ the usual Lanczos iteration with an appropriate nuclear Hamiltonian, with the twist that our start vector is taken to be a very highly excited state.  

Using the shell-model code Oxbash \cite{oxbash}, we performed a full sd-shell calculation of $^{28}$Si using the USDB Hamiltonian \cite{br:2006} (closed $^{16}$O core with 12 valence nucleons in the 1d and 2s shells).  We performed a full fp-shell calculation of $^{47}$Ti using the GPFX1 Hamiltonian \cite{hobm:2004} (closed $^{40}$Ca core with 7 valence nucleons in the 1f and 2p shells).  Finally, we performed a truncated fp-shell calculation for $^{56}$Fe using the GPFX1 Hamiltonian, only allowing up to 2 valence protons and up to a total of 4 valence nucleons to occupy single-particle states above the zero-order (no residual interaction) ground state configuration. Some of the $sd$ calculations were carried out with the NuShellX code \cite{nushellx}.

When experimental Gamow-Teller beta-decay strengths are compared to the results obtained from calculations in the $sd$ and $pf$ model spaces it is observed that experimental strengths are uniformly reduced relative to theory by a factor of 0.5-0.6 \cite{bw:1988,mpcz:1996}.  This ``quenching" is mainly due to second-order configuration mixing induced by the short-ranged part of the tensor interaction \cite{asbh:1987,towner:1987}.  We assume that the same quenching applies to transitions from the excited states and we use a reduction factor of $r=0.50$.

\begin{figure}[here]
\centering
\includegraphics[scale=.68]{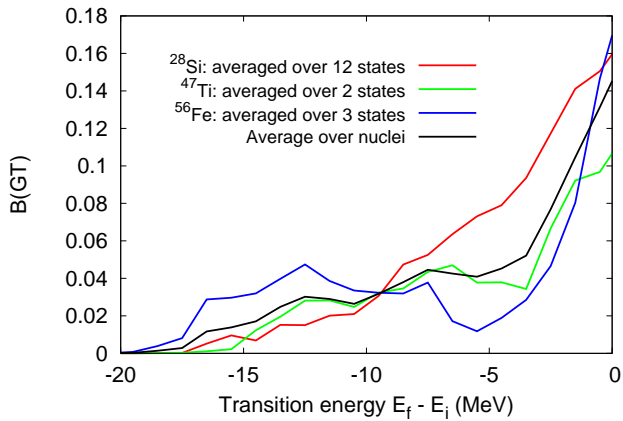}
\caption{Neutrino-pair emission transition strength for $^{28}$Si, $^{47}$Ti, and $^{56}$Fe from initial states at 27.6 MeV excitation shown as functions of the difference between initial excitation energy E$_i$ and final excitation energy E$_f$.}
\label{fig:Nucstrength}
\end{figure}

We examined transition strengths and energy loss rates over a range of excitation energies from 0 to 40 MeV in $^{28}$Si and $^{56}$Fe and at 23 and 27.6 MeV excitation in $^{47}$Ti.  The strength distributions for all three nuclei at 27.6 MeV excitation are shown in Fig.~\ref{fig:Nucstrength} along with the distribution obtained by averaging the strength as a function of transition energy over all three nuclei.  While the details of the shapes of the strength distributions vary between nuclei, the essential feature of a central peak with a long tail out to transition energies of 15 or 20 MeV is consistent.

\begin{figure}[here]
\centering
\includegraphics[scale=.68]{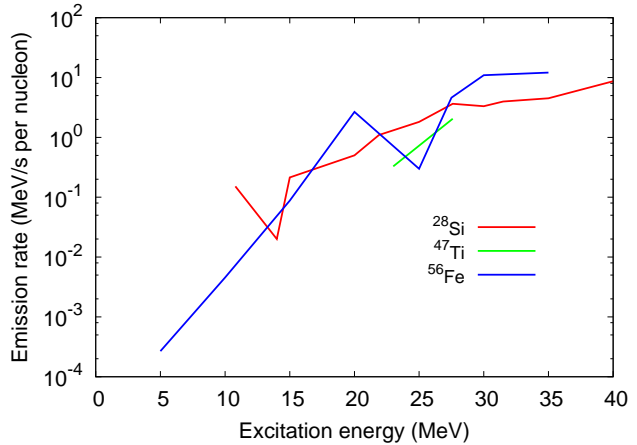}
\caption{Energy loss by neutrino-pair emission for $^{28}$Si, $^{47}$Ti, and $^{56}$Fe shown as functions of excitation energy.}
\label{fig:rates_excitation}
\end{figure}

To obtain energy emission rates per nucleon as functions of excitation energy, shown in Fig.~\ref{fig:rates_excitation}, we applied Eq.~\ref{eq:ratetotal} to each nucleus, then divided by A.  With an eye toward extension to large nuclei, the key observation is that the energy loss rate {\it per nucleon} is strongly dependent on excitation energy, but nearly independent of nucleus, despite the considerable differences in the models used for each nucleus.

\begin{figure}
\centering
\includegraphics[scale=.68]{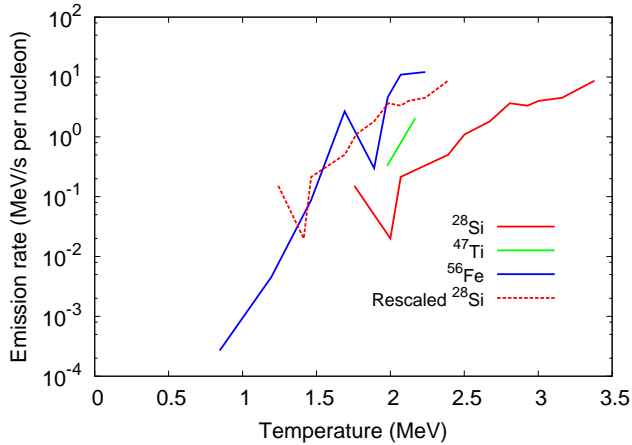}
\caption{Energy loss by neutrino-pair emission for $^{28}$Si, $^{47}$Ti, and $^{56}$Fe shown as functions of temperature.  The dashed line shows the $^{28}$Si result with the temperature scaled by a factor of $\left( \frac{28}{56}\right)^{1/2}$, allowing the  $^{28}$Si and $^{56}$Fe results to be compared directly at equivalent excitation energies, as in Fig.~\ref{fig:rates_excitation}.}
\label{fig:rates_temperature}
\end{figure}

To find temperature as a function of mean excitation energy and nucleus, we invert Eq.~\ref{eq:bethe}.  This, along with the approximation in Eq.~\ref{eq:approx}, gives emission rate per nucleon as a function of temperature, shown in Fig.~\ref{fig:rates_temperature}.  Also shown is the result for $^{28}$Si, but with the temperature computed from excitation energy as though it had the same mass number as $^{56}$Fe.  As can be seen from Eq.~\ref{eq:bethe}, this amounts to scaling the temperature of $^{28}$Si by a factor of $\left( \frac{28}{56} \right)^{1/2}$.  This allows us to compare the $^{28}$Si (our most realistic model) and $^{56}$Fe (our most astrophysically relevant nucleus) results directly as functions of temperature: the comparison of the scaled results in Fig.~\ref{fig:rates_temperature} is equivalent to what is shown in Fig.~\ref{fig:rates_excitation}.  We use this method of temperature scaling extensively throughout this paper; it will be indicated in each case.

\section{Results}

\begin{figure}
\centering
\includegraphics[scale=.65]{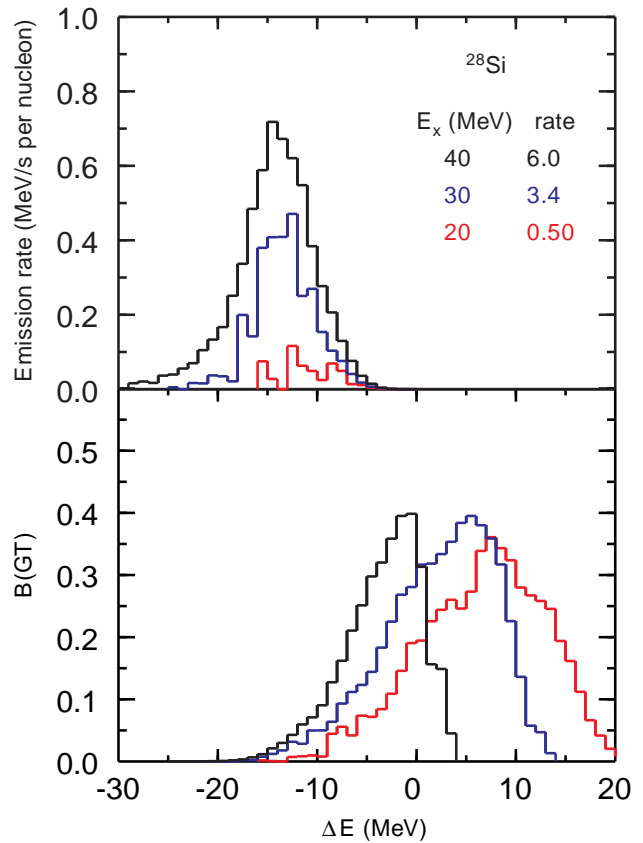}
\caption{Bottom: Transition strengths for $^{28}$Si from states at 20, 30, and 40 {\rm MeV} as functions of transition energy.  Top: Energy emission rates per nucleus via neutrino pairs for $^{28}$Si as functions of transition energy, {\it i.e.}, spectra of emitted neutrino pairs.}
\label{fig:si28}
\end{figure}

We found that the weak strength in transitions from highly excited initial states is spread significantly in energy.  The $^{28}$Si results for the distribution of axial vector strength (B(GT): squared matrix element) with transition energy and the corresponding neutrino-pair energy spectrum are shown in the lower and upper panels, respectively, of Fig.~\ref{fig:si28}. This plot shows both \lq\lq up\rq\rq\ transition strength, corresponding to positive values of $E_f-E_i$, and \lq\lq down\rq\rq\ strength with negative values of $E_f-E_i$ appropriate for de-excitation into neutrino pairs. At 20 and 30 {\rm MeV} excitation, there is obviously more strength in the up than in the down channel. However, the up-strength distributions for both 30 and 40 {\rm MeV}  are not complete (artificially cut off) due to a truncation in the calculation at 10,000 Lanczos iterations. We obtained these distributions by averaging over ten states with $J_{i}=5$ near each indicated initial excitation energy. $J_{i}=5$ was chosen because the $sd$ shell $(2J+1)$ state density peaks at this spin value.

According to Eqs.~\ref{eq:btz} and \ref{eq:ratetotal}, the strength distribution as a function of transition energy can be multiplied by six powers of the transition energy and a constant to give the contribution of a given transition energy to the overall neutrino-pair energy emission rate, {\it i.e.}, the neutrino-pair energy spectrum.  Although the actual strength distribution is skewed toward lower energy transitions, weighting with six powers of transition energy clearly favors larger transitions.  Summing over transition energy ({\it e.g.,} Eq.~\ref{eq:ratetotal}) gives the total neutrino-pair energy emission rate per baryon.  Table \ref{table:rates} shows total energy loss rates per baryon computed in this fashion for several nuclei over a range of excitation energies and temperatures (rescaled as in Fig. \ref{fig:rates_temperature}).

\begin{figure}
\centering
\includegraphics[scale=.64]{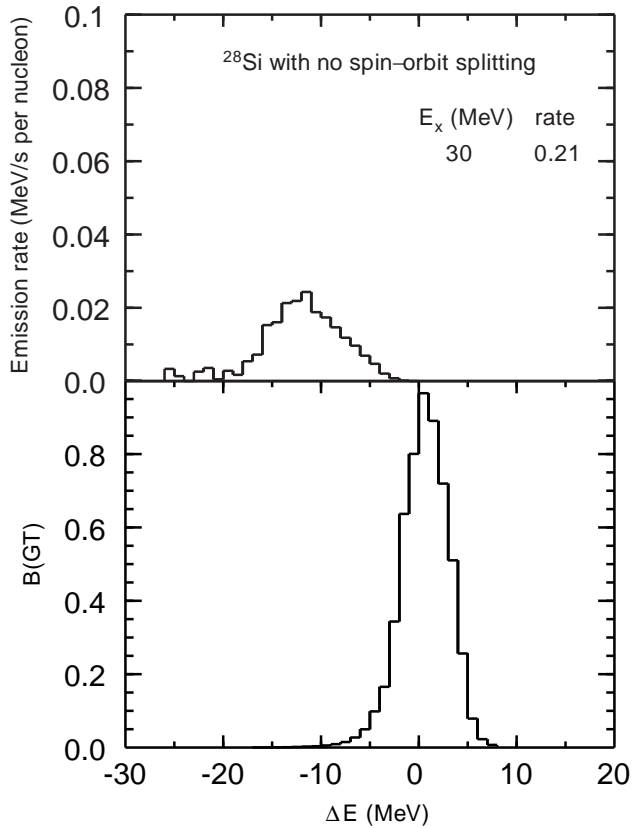}
\caption{$^{28}$Si strength and neutrino pair spectrum at 30 {\rm MeV} excitation computed with spin-orbit splitting in the nuclear Hamiltonian set to zero.  The strong central peak, lack of wings, and concomitant drastic reduction in neutrino production confirm the role of spin-orbit splitting and spin-flip transitions in producing the high rates computed in our more realistic models.}
\label{fig:si28ns}
\end{figure}

Our calculations show that the central peaks in the Fig.~\ref{fig:si28} strength distributions stem primarily from lateral (no spin flip) transitions. Such
transitions do not change the single-particle energy of the  transitioning nucleon. The wings of these strength spectra at larger transition energy come mostly from nucleon spin-flip transitions.  Because configurations that result from a spin flip have a lower zero-order energy as a result of particle-hole repulsion and spin-orbit splitting, they tend to be more readily mixed down to lower excitation energy than their counterparts stemming from no-spin-flip, lateral transitions.  This became abundantly clear when we computed $^{28}$Si with no spin-orbit splitting in the nuclear Hamiltonian, treating the $\ell+1/2$ and $\ell-1/2$ single-particle states as though they have the same energy.  The results of this computation are shown in Fig.~\ref{fig:si28ns}, where it is readily seen that all of the strength is concentrated in a central peak, drastically reducing the rate of neutrino pair production. Since the spin-orbit splitting is due to the nuclear surface, we conclude that the Gamow-Teller down-strength is greatly enhanced when the baryons are confined to nuclei.  

\begin{table}[here]
\begin{tabular}{ | c | c | c | c | c | }
\hline
Nucleus & $J_i$ & Excitation & Rate & T (A=56)\\
\hline
$^{28}$Si & 1-4 & 10.8 & 0.30 & 1.2\\
& 0, 2-5 & 14.0 & 0.02 & 1.4\\
& 0-5 & 15.0 & 0.22 & 1.5\\
& 5 & 20 & 0.50 & 1.7\\
& 0-5 & 20.0 & 0.50 & 1.7\\
& 0, 2-5 & 21.9 & 1.1 & 1.8\\
& 5 & 25 & 1.8 & 1.9\\
& 1-5 & 25.0 & 1.8 & 1.9\\
& 0-5 & 27.6 & 3.6 & 2.0\\
& 5 & 30 & 3.4 & 2.1\\
& 0-3, 5& 30.0 & 3.3 & 2.1\\
& 1-5 & 31.5 & 4.0 & 2.1\\
& 5 & 35 & 4.5 & 2.2\\
& 5 & 40 & 6.0 & 2.4\\
& 0-5 & 40.0 & 8.6 & 2.4\\ \hline
$^{28}$Si (no SO) & 5 & 30 & 0.21 & 2.1\\
$^{29}$Si & 11/2 & 30 & 4.6 & 2.1\\
$^{28}$P & 5 & 30 & 9.7 & 2.1\\ \hline
$^{47}$Ti & 3/2 & 23.0 & 0.33* & 1.8\\
& 3/2, 5/2& 27.6 & 2.0* & 2.0\\ \hline
$^{56}$Fe & 2 & 10.0 & 0.01* & 1.2\\
& 0 & 15.0 & 0.09* & 1.5\\
& 1 & 20.0 & 2.7* & 1.7\\
& 2 & 25.0 & 0.30* & 1.9\\
& 0, 1 & 27.6 & 4.6 & 2.0\\
& 0 & 30.0 & 10.9* & 2.1\\
& 4 & 35.0 & 12.1* & 2.2\\
\hline
\end{tabular}
\caption{Energy loss rate (MeV/s/baryon) for various nuclei as functions of excitation energy (MeV) and the corresponding (rescaled to A=56) temperature (MeV).  The unstarred rates were computed by averaging over several (5 to 14) initial states at the indicated excitation energy; starred rates were computed from 1 or 2 states. The angular momenta of all initial states are indicated. The entry $^{28}$Si (no SO) was computed by neglecting spin-orbit splitting in the nuclear Hamiltonian.}
\label{table:rates}
\end{table}

\section{Discussion and Conclusions}

We have found that transitions between spin-orbit partners account for the bulk of the spread to lower excitation energies (hence, larger $\Delta$E) of the Gamow-Teller strength in this channel.  That actually bodes well for any attempt to use the nuclear systematics of lighter nuclei like $^{28}$Si and $^{56}$Fe to effect an extrapolation of neutrino pair emission mechanisms to the higher mass nuclei of most interest in stellar collapse.  This is because the spin-orbit splitting is relatively constant across nuclear mass in the range over which we are interested \cite{bm:1969}.  Particle-hole repulsion probably plays a lesser role than spin-orbit splitting in pushing strength to larger $\Delta$E. 
Interestingly, the particle-hole repulsion we find in our shell-model calculations may have a direct analog in the bulk matter renormalization of the energetics of weak interaction processes as found in Ref.~\cite{Shen:2012fk}.

There are three obvious effects of skewing the strength distributions to higher transition energy $\Delta E$. These follow from the simple fact that the neutrino-pair energy emission rate derived from the strength function is weighted by six powers of $\Delta E$. First, more strength at higher $\Delta E$ generally means faster neutrino-pair emission rates. In turn, this means more energy will be pumped into neutrino pairs by this process. Second, the neutrinos and antineutrinos produced by this process will have higher energies on average. This brings up an obvious question: will the neutrino pair energies now be so high ($ > 10\,{\rm MeV}$) that they are more readily trapped?  Third, more configurations in play and more configuration mixing at higher excitation energy will make the transition strength and the neutrino-pair emission rate more sensitive to temperature.

Higher temperatures and higher excitation energies bring up an issue which is unresolved in our work.  When nucleons are promoted into the next-higher oscillator level, how is the Gamow-Teller strength affected? For example, for $^{28}$Si the actual level density just above about 10 MeV will start to be dominated by one-nucleon excitations into and out of the $sd$ shell (negative-parity states). The Gamow-Teller down-strength for these states will be reduced since they must go to negative-parity final states that only exist at a higher energy than exist for positive parity. Starting at 20 {\rm MeV}, more positive parity states can be made from two nucleons excited into and out of the $sd$ shell. Does this result in a lower overall amount of strength?  Or is this loss of strength compensated by more transitions between spin-orbit partners in the higher energy shell?  In general, it may seem reasonable that the higher the temperature and excitation energy, the ``looser" the nucleus and the more transitions are unblocked \cite{fuller:1982}.  However, this trend could be thwarted by the actual behavior of the level density in the shell model.  Certainly, model space truncation could contribute to this if, for example, not all spin-flip transition channels in the higher oscillator level are included in the calculation. This issue may or may not complicate extrapolation of our trends in weak strength energy distributions to higher mass nuclei and requires further investigation.

\begin{figure}[here]
\centering
\includegraphics[scale=.68]{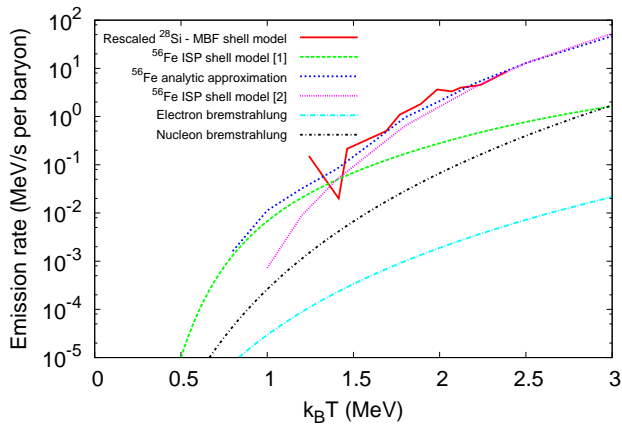}
\caption{Energy per baryon emitted in neutrino pairs as a function of temperature.  Solid: $^{28}$Si as computed in this paper.  Dashed: $^{56}$Fe from Fuller \& Meyer analytic approximation \cite{fm:1991}.  Dotted: $^{56}$Fe from Fuller \& Meyer independent single-particle shell-model calculation \cite{fm:1991}.  Fine dotted: $^{56}$Fe from Kolb \& Mazurek \cite{km:1979}. Long dash-dotted: Electron bremsstrahlung into neutrino pairs for $\rho_{12}=1$ from Dicus et al \cite{dkst:1976}.  Short dash-dotted: Nucleon bremsstrahlung into neutrino pairs at nuclear matter density from Friman \& Maxwell \cite{fm:1979}.}
\label{fig:ratecomparison}
\end{figure}

Fig.~\ref{fig:ratecomparison} shows a comparison between our shell-model calculations of the neutrino-pair energy emission rate per baryon for $^{28}$Si as a function of (rescaled) temperature, other estimates for this rate in $^{56}$Fe as a function of temperature, the neutrino pair emission rate for electron bremsstrahlung as a function of temperature assuming a density $\rho_{12}=1$ \cite{dkst:1976}, and the neutrino pair emission rate for nucleon bremsstrahlung in nuclear matter as a function of temperature \cite{fm:1979}. We include the nucleon bremsstrahlung result to show that collectivity within the nucleus enhances the emission rate relative to bulk nuclear matter.  We found that in the temperature regime of interest, our calculations can yield neutrino-pair energy emission rates that equal or exceed earlier estimates \cite{fm:1991,km:1979} at all temperatures.  As a general rule, our estimates of these rates are approximately three orders of magnitude faster than neutrino pair production from electron bremsstrahlung.

Clearly, de-excitation of nuclei is the dominant contributor of relatively low-energy neutrino pairs under these conditions.  Moreover, the rates presented here are lower bounds on the actual neutrino-pair production rates, particularly at temperatures between 1 and 1.5 {\rm MeV}.  The small number of nucleons in a $^{28}$Si nucleus gives a relatively low density of states at the temperatures of interest.  As a consequence, there are few lower-lying states to transition to, which reduces the total transition rate.  Indeed, the apparent decrease in emission rate of $^{28}$Si at a (rescaled) temperature of 1.4 {\rm MeV} is a consequence of the simple fact that no states near 14 {\rm MeV} excitation have transitions with energies greater than those available to states near 10.8 {\rm MeV} transition.

{\it If} it turns out that the neutrinos produced by pair de-excitation have low enough energies to escape the pre-supernova star, then this process likely acts as a thermostat for the collapsing core. In this limit, as the core heated up, more neutrino pairs would be produced and escape, carrying away entropy, and perhaps keeping the core temperature near $T=1\,{\rm MeV}$ to $1.5\,{\rm MeV}$.

However, our calculations may be suggesting that the neutrino pairs produced by de-excitation at higher temperature are so energetic that they do not escape. Though the thermostat effect will be disabled in this case, lower to intermediate energy neutrino phase space will be filled more quickly by this process, and the core will approach beta equilibrium sooner. This effect would tend to block electron capture and neutronization sooner also, but against this neutrino-nucleus and neutrino-electron down scattering will tend to heat the system, adding entropy, implying faster electron capture through more free protons and nuclear thermal unblocking \cite{fuller:1982} and, therefore, a lower $Y_e$, a smaller homologous core, and a concomitantly lower initial bounce shock energy.

The issue of higher energy neutrino-pairs is complicated further when considering the effects of dynamics and flavor.  The energy is shared between a neutrino and an anti-neutrino, and the energy is small compared to the mass of the nucleus.  As a consequence, the energy can be shared unequally between the two neutrinos, with the nucleus absorbing whatever momentum is needed to satisfy conservation.  So a low-energy partner to a high-energy neutrino could escape.  Furthermore, this process is flavor blind: it produces neutrinos of all flavors at equal rates.  Because high energy neutrinos within the collapsing core are produced primarily from electron capture, there is plenty of available phase space for nuclear de-excitation to produce a high energy electron anti-neutrino or a high energy neutrino or anti-neutrino of mu or tau flavor with a low energy partner that easily escapes.  The resultant asymmetry between electron flavored neutrinos and anti-neutrinos could impact lepton number within the core, as it is more likely to produce a high energy electron anti-neutrino with a low energy partner than the other way around.

Another effect of the larger $\Delta E$ values suggested by our shell-model calculations may be an enhancement in the plasmon-mediated neutrino-pair nuclear de-excitation process pointed out by Horowitz in Ref.~\cite{Horowitz:1992lr}. The matrix element for the first forbidden vector channel considered in Ref.~\cite{Horowitz:1992lr} is $\propto \langle f\vert q \cdot T_z \vert i\rangle$, where $q \sim \Delta E$ is the momentum transfer. This first forbidden channel is in general cut down by a geometric factor, $ {\left( qR\right)}^2 \sim 1/16$,  which is the square of the ratio of the nuclear radius $R$ to the inverse momentum transfer. Our larger values of $\Delta E$ should give a smaller reduction, increasing the overall rate of nuclear de-excitation into pairs.

Though our shell-model calculations are only a beginning, they do suggest that nuclear de-excitation into neutrino pairs is likely the dominant source of low to intermediate energy neutrino pairs in stellar collapse. Our calculations suggest a spin-orbit splitting-induced increase in the rate of this process and a steepening of the temperature dependence of this rate. These calculations also suggest, however, that the neutrinos produced in this process are more energetic and may be trapped. Only inclusion in a full core collapse neutrino transport simulation could reveal what role this process plays in core collapse supernova explosions.

\begin{acknowledgments}
This work was supported in part by NSF grant PHY-09-70064 at UCSD and NSF grant PHY-1068217 at MSU.  
We would also like to acknowledge support from the LANL Topical Collaboration and UCOP, and support of the Michigan State University High Performance Computing Center and the Institute for Cyber-Enabled Research. We would like to thank J. Carlson, J. Cherry, and S. Reddy for helpful discussions, and A. Y. Shih for assistance in creating Figs.~\ref{fig:feyn} \& \ref{fig:ei-ef}.
\end{acknowledgments}

%\bibliography{/Users/gfuller/XBibTeX-references/allref}
\bibliography{references.bib}

\end{document}